\documentclass[12pt]{article}
\input epsf.sty
\def\be{\begin{equation}} \def\ee{\end{equation}}
\def\bea{\begin{eqnarray}} \def\eea{\end{eqnarray}} \def\ba{\begin{array}}
\def\ea{\end{array}} \def\ben{\begin{enumerate}} \def\een{\end{enumerate}}

\newcommand{\eqn}[1]{(\ref{#1})}

\newcommand{\hepth}[1]{{\tt hep-th/{#1}}}

\def\l{\lambda}
\def\m{\mu}

\def\n{\nu}

\def\ov{\over}
\def\br{\nonumber\\}

\def\g{\gamma}
\def\G{\Gamma}
\def\Tr{{\rm Tr}}

\begin{document}
{}~
\hfill\vbox{\hbox{hep-th/yymm.nnnn} \hbox{\today}}\break

\vskip 3.5cm
\centerline{\large \bf
More on the {\it scalar-tensor} B-F theory }
\vskip .5cm

\vspace*{.5cm}

\centerline{  Harvendra Singh}

\vspace*{.25cm}
\centerline{ \it  Theory Group, Saha Institute of Nuclear Physics} 
\centerline{ \it  1/AF, Bidhannagar, Kolkata 700064, India}
\vspace*{.25cm}

\vspace*{.5cm}

\vskip.5cm
\centerline{E-mail: h.singh [AT] saha.ac.in }

\vskip1cm
\centerline{\bf Abstract} \bigskip
This work is based on an earlier proposal \cite{hs} that the membrane 
B-F theory consists of 
 matter fields alongwith Chern-Simons  fields as well as
the auxiliary pairs of scalar and tensor 
fields. We especially discuss the supersymmetry aspects of such a 
membrane theory.  It is concluded that the 
theory  possesses
maximal supersymmetry and it is related to the L-BLG theory via a 
field map. We obtain  fuzzy-sphere 
solution and  corresponding  tensor field configuration is 
given.
  
\vfill \eject

\baselineskip=16.2pt

\section{Introduction}
Recent
advances in 3-dimensional matter Chern-Simons field theories
 have led to some
interesting proposals for the superconformal field theory
describing  super-membranes living in eleven spacetime dimensions.
Amongst these, the Bagger-Lambert-Gustavsson (BLG) theory  
has ${\cal N}=8$ superconformal invariance but so far this theory has been 
constructed
 explicitly for a compact $SO(4)$ gauge group only \cite{bl,Gustav}.
While allowing for  noncompact (Lorentzian) tri-Lie-algebras, the BLG 
framework has been
extended further to admit full $SU(N)$ gauge symmetry 
\cite{gomis,verlinde}. 
But these latter ones, also known as L-BLG theories, have
ghost fields in their spectrum. Once the ghost fields are eliminated 
through gauging procedure  
the theory eventually reduces to the $SU(N)$  super Yang-mills theory
\cite{Bandres:2008kj}. On the other hand, another interesting class of 
matter Chern-Simons 
theories,  known as ABJM  theories 
\cite{abjm}, 
however are  based
on ordinary Lie-algebras involving bi-product gauge groups. The ABJM 
theory  admits
${\cal N}=6$,
$U(N)_k\times U(N)_{-k}$ superconformal
symmetry, and these are conjectured to be dual to  M-theory compactified 
on $AdS_4\times
S^7/Z_k$ spacetime, with arbitrary level $k>2$. Only when $k=1,2$, the 
theory supposedly becomes a
maximally supersymmetric theory. The $AdS_4$ 
geometry arises in the near horizon limit when 
$N$ M2-branes are placed at the singularity in an 8-dimensional 
orbifold space $C_4/Z_k$ \cite{abjm}. 
\footnote{Specifically, 
 M2-brane solutions on a `resolved' 
$C_4/Z_4 $ space
and corresponding Chern-Simons level flow in ABJM theory 
have been studied in 
\cite{chethan}. It is shown that the M2-brane solutions are smooth 
when branes are placed on the resolution.}
The  two  theories BLG and ABJM complement each other but the theories 
have very distinct field theoretic
 structures though. 
Particularly in the context of L-BLG theories, it has become imperative 
to  explore the fundamental importance of tri-algebras in a
membrane theory.\footnote{The primary motivation 
for studying
3-algebras in membrane theory context  arose from the work 
\cite{basu}}  
Along this direction there have been 
works where the maps between L-BLG and ABJM theories are explored in 
detail 
\cite{tseytlin}. \footnote{ 
Also specially see \cite{sudipto} for a divergent study of  Jordan 
algebras  in the BLG framework.}    
 Particularly, our motivation in this paper shall be not to emphasize 
on the tri-algebra aspects,
instead we  simply  try to work with ordinary Lie-algebra 
 so long as it is possible.
 
Following
various works \cite{mukhi,bobby,verlinde} on 
 B-F (Chern-Simons) and L-BLG Lagrangians, in a recent paper
\cite{hs} we showed 
that one can construct membrane B-F  
theories simply using ordinary Lie algebra. The crucial 
difference had been that unlike in the L-BLG 
construction which relies
upon the introduction of pair of propagating (ghost) fields 
 $(X^{+},X^{-})$,
 our 
construction  instead requires  introduction of pairs of
scalar and tensor 
 fields. The tensor fields are introduced through  `BdGPT' like field 
duality as in the Romans theory, it is discussed in the Appendix here. 
Interestingly, 
these dual-pairs of scalar and tensor fields 
remain non-propagating in the action just like the Chern-Simons 
fields. 
In the present paper we work within the axiom that 
the $SU(N)$ membrane theory has 
fundamental propagating scalar fields along with auxiliary B-F 
 gauge fields and auxiliary scalar-tensor 
fields and their superpartners. Incidentally, eight 2-rank tensor fields 
$C_{\m\n}^I$ appear only through their topological coupling with dual 
scalar fields 
$\eta^I$ as 
\be\label{coupl1}
\int \eta^I  d C^I_{(2)} \ee
The vev   $<\eta^I>$ eventually gets related to the 
coupling constant of the 3D super-Yang-Mills theory. Thus the strength of 
the coupling 
constant determines the presence of tensor fields in the membrane 
BF theory. 
If the coupling  vanishes in the vacuum so also the tensor fields. 
The 
presence of $C_{(2)}^I$ perhaps may also be motivated from the 
membrane boundary point of view. 
An open-membrane is a 2-dimensional extended 
object and its boundary (taking for example M2-brane ending on M5-brane) is 
essentially an extended string-like  configuration which 
can inherit a fundamental tensor field 
$C_{\m\n}$. Such one-dimensional extended solitonic excitations would of 
course live in the 
world-volume theory of M5-branes. This is essentially the 
argument also used by Basu and Harvey \cite{basu} in order to propose 
tri-algebras.
We do know there are  solitonic string solutions on M5-branes
with self-dual 
3-form tensor fields ${\bf C}_{\m\n\l}$ along its world-volume 
\cite{howe}. 
So when M2-branes end on 
 M5-brane, by gauge symmetry argument, we should define a 
gauge invariant field 
strength $({\bf C}_{\m\n\l}<\eta^I>-\partial_{[\m} C_{\n\l ]}^I)$ on 
the M5-brane.\footnote{It is not clear whether a formulation of 
$M5$-brane theory exists with this kind of field structure. However, 
an important covariant formulation of M5-brane  theory with 
field structure  $({\bf C}_{\m\n\l}-\partial_{[\m} C_{\n\l ]})$ 
and with an auxiliary scalar field has been studied 
in \cite{tonin}. } 
While from M2-brane point of view  
the membrane having nontrivial boundary  configuration 
should correspondingly  include a tensor field 
$C_{\m\n}^I$ (non-propagating) in its world-volume theory, such as 
the coupling in \eqn{coupl1}. The above argument 
appears similar in spirit to the case when open-strings end on 
D$p$-branes. 
The string end-points 
are charged with gauge (Chan-Paton) fields 
which give rise to the topological (gauge) coupling 
\be
g_s\int_{\partial\Sigma} 
{\cal 
A}_{(1)} \ee in the open-string 
world-sheet theory and also give rise to a dynamical gauge theory on 
the D$p$-brane itself.

Our goal in this paper is to extend our earlier work \cite{hs} 
and 
specially discuss the supersymmetry aspects of  the B-F 
theory with  tensor fields. We shall show that the 
theory has a maximal supersymmetry. 
We also discuss supersymmetric  
solutions, particularly 
the fuzzy-sphere solution, and obtain
corresponding nontrivial tensor field  responsible 
for this solution.
We also comment on the  equivalence 
between our ordinary Lie-algebra theory and the  tri-Lie-algebra 
based L-BLG theories.  

The paper is organised as follows. In the section-2 we review the main 
aspects and symmetries of the membrane B-F action. In the section-3 we 
provide a supersymmetric completion of this theory. We then discuss the 
equivalence between our work and the L-BLG frame work. The section-4 
deals with the supersymmetric fuzzy $S^2$ solution and we discuss the 
hidden aspects of the shift-symmetry. The conclusions are given in 
section-5.

\section{Review: STBF theory}
The bosonic part of the membrane B-F action proposed in \cite{hs}, or 
more appropriately called  scalar-tensor B-F (STBF) action 
here,  is given by
  \bea\label{MCS}
&&S_{STBF}=\int d^3x \bigg[ \Tr(
 -{1\ov2}(D^\mu X^I-\eta^I B^\mu)^2 + {1\ov2}
\epsilon^{\mu\nu\lambda} B_\mu F_{\nu\lambda}-U(\eta,X))
\br &&~~~~~~~~~~~~~ -{1\ov2}\epsilon^{\mu\nu\lambda} C_{\mu\nu}^I 
\partial_\lambda\eta^I \bigg]
\eea
where 
\bea
&& D_\mu X^I=\partial_\m X^I - [A_\m,X^I]\ , ~~~~~\br
&&V_{IJK}=\eta_{[I} 
X_{JK]}=\eta_I X_{JK}+{\rm cyclic~ permutations~of~ indices}\ ,\br
&&U={1\over 2.3!}(V_{IJK})^2 \ .
\eea
Here $X_{JK}=[X_J,X_K]$ is the Lie bracket. The $X^I$'s $(I=1,\cdots,8)$ 
are the scalars 
while $B_\mu$ and $A_\mu$ are the Chern-Simons gauge fields. All fields 
are in the adjoint 
of $U(N)$ except the scalars $\eta^I$ and the tensors $C_{\mu\nu}^I$ 
which are singlets. Note 
that tensor fields appear only as Lagrange multipliers.

Various equations of motion are: namely the $X^I$ equation
\be
\partial_\mu(D^\mu X^I-\eta^I B^\mu)-
[A_\mu,(D^\mu X^I-\eta^I B^\mu)]-\partial_{X^I}U=0\ ,
\ee
the $B_\mu$ equation (or the dNS-duality relation \cite{nicolai}) 
\be\label{eqxi}
{1\over 2! }\epsilon^{\mu\nu\lambda} 
F_{\nu\lambda}
=-(D^\mu X^I-\eta^I B^\mu)\eta^I\ ,
\ee
the $ C^I_{\m\n}$ equation
\be\label{new1}
\partial_\l\eta^I=0\ ,
\ee
and the $\eta^I$ equation
\be\label{new2}
 \Tr((D^\mu X^I-\eta^I B^\mu)B_\mu - {1\over2} V^{IJK}X_{JK}) 
+{1\over 2}\epsilon^{\m\n\l} \partial_\m C_{\n\l}^I=0 \ .
\ee
Thus $\eta^I$'s are constants in a given vacuum. The 
equation \eqn{new2} does relate 
$\eta^I$ with its dual field $ C_{\n\l}^I$ and should be taken as 
the Hodge-duality (BdGPT) relation, literally in the same sense as in 
Romans' 
type IIA supergravity theory, see 
Appendix for details. In this way, the fields $\eta^I$ and $C_{\m\n}^I$ 
form  dual-pair of fields. Note that, in this form of the action the 
scalar-tensors 
and the Chern-Simons $(B_\m,A_\m)$ fields are at the same footing. 
They all are 
auxiliary fields. So it will be 
more appropriate to call the above theory as scalar-tensor B-F or simply 
STBF membrane theory.
 There are however
no free parameters in the theory. 

The action  has an scale invariance 
\bea
&& x_\mu \to a^{-1} x_\mu, ~~~X^I\to a^{1/2} X^I, ~~~(B_\mu, A_\mu)\to (a 
B_\mu, a A_\mu),\br
&& (\eta^I, C^I_{\mu\nu})\to (a^{1/2} \eta^I, a^{3/2}C^I_{\mu\nu})
\eea
where $a$ is an arbitrary  scale parameter.

The gauge symmetry of the action is
\bea
&&X^I\to U^{-1}X^I U,~~~
A_\mu\to U^{-1}A_\mu U -U^{-1}\partial_\mu U, \br
&&B_\mu\to U^{-1}B_\mu U,
\eea
where $U\in U(N)$. Note that the $B_\mu$ field 
transforms as an 
adjoint field like $X^I$ but distinctly as compared 
to the gauge field $A_\mu$. The noncompact shift symmetry under which 
$X^I$ transforms as $X^I\to X^I+\eta^I M$, where $M$ is arbitrary 
\cite{verlinde},  
is not the symmetry of the action \eqn{MCS} because $\eta^I$'s are not 
constant. 
However, it remains a symmetry in a given vacuum, that is when 
$<\eta^I>$ become constant. In order 
to recover 
the shift symmetry in the action itself we will need to add compensating 
terms, as we discuss it
next along with supersymmetry.

 Note that, in the vacuum we shall have coupling constants $g^I$ 
which gets rotated under $SO(8)$. The identification of these couplings 
goes as 
\be
g^I=~ <\eta^I(x)>, ~~~~g^Ig^I=(g_{YM})^2
\ee
where $g_{YM}$ is the Yang-Mills coupling constant in the D2-brane 
gauge theory. The B-F action \eqn{MCS} has a new $U(1)$ invariance 
under
\be
  C_{(2)}^I \to  C_{(2)}^I + d\alpha_{(1)}^I \ ,
\ee 
where $\alpha_{(1)}^I$ are arbitrary 1-forms.

The dNS relation and the BdGPT relation in eq.\eqn{new2} can be combined 
to give an  
identity
\be\label{eqquan}
 \Tr({1\over 2! }\epsilon^{\mu\nu\lambda} 
F_{\nu\lambda}B_\mu + U) 
={1\over 2}\eta^I\epsilon^{\m\n\l} \partial_\m C_{\n\l}^I
\ee
This is an useful relation. It implies that there can always be a 
nontrivial tensor field in the vacuum whenever the gauge fields are 
nontrivial or 
when there is a nontrivial potential. Particularly, in an Abelian theory 
$U=0$, the gauge fields have to be present for tensor fields to 
be nontrivial. We shall give an example  where tensor fields 
are nontrivial. 

In summary, the B-F theory has 
actually  two sets of 
pair of fields, the dNS adjoints $(B_\mu, A_\mu)$ and 
the BdGPT singlets $(\eta^I,C^I_{\m\n})$. The 
introduction of 
these pairs  has  helped in bringing YM theory into the B-F Lagrangian 
form which has explicit
$SO(8)$ global invariance and $U(N)$ gauge symmetry.
In the  work \cite{hs}, it was left to determine what is the actual 
supersymmetry content of this scalar-tensor B-F theory as
only bosonic part of the Lagrangian was presented there. Here we determine 
the 
full ${\cal N}=8$ supersymmetry content of the theory.

\section{Supersymmetry}
\subsection{ The U(1) case}
To help the task we 
discuss the Abelian case first as the potential vanishes in 
this case. 
The  
 Lagrangian for a single membrane can be obtained from the 
above  STBF action  and it is 
\bea\label{MCS1}
&&S_{U(1)}=\int d^3x \left(
 -{1\ov2}(\partial^\mu X^I-\eta^I B^\mu)^2 + {1\ov2}
\epsilon^{\mu\nu\lambda}  B_\mu F_{\nu\lambda} 
-{1\ov2}\epsilon^{\mu\nu\lambda} C_{\mu\nu}^I 
\partial_\lambda\eta^I 
-\partial_\lambda\eta^I (B^\l X^I)\right)
\br &&~~~~~~~~~~ 
\eea
Note that an additional term $-\partial_\lambda\eta^I 
(B^\lambda X^I)
$ has  been added to the action \eqn{MCS1} so that it now has a 
 shift  
 (Stueckelberg)
symmetry
\bea\label{shift1}
\delta_1 B_\mu=\partial_\mu f, ~~~
\delta_1 X^I=\eta^I f, ~~~
\delta_1 C_{\mu\nu}^I=\epsilon_{\mu\nu\lambda}\partial^\lambda (f X^I), 
\eea
in addition to the Abelian gauge invariance under the variation
\be
\delta_2 A_\mu=\partial_\mu \lambda .
\ee

With the information about the supersymmetric scalar-tensor 
topological action
given in Appendix, 
\be
 S_{ST}=-\int d^3x ({1\ov2}\epsilon^{\mu\nu\lambda} C_{\mu\nu}^I 
\partial_\lambda\eta^I
+i\bar \chi^A\zeta^A) ,
\ee
we  find that a 
supersymmetrised Abelian STBF action is 
\bea\label{MCS1a}
&&S_{U(1)}=\int d^3x \, (
 -{1\ov2}(\partial^\mu X^I-\eta^I B^\mu)^2 + {1\ov2}
\epsilon^{\mu\nu\lambda} B_\mu F_{\nu\lambda}\br
&&~~~~~~~~-\partial_\lambda\eta^I (B^\l X^I)
+ {i\over 2} \bar\psi
\not\!\partial\psi +i\bar\chi \not\!B \psi
+S_{ST})
\eea
where $\psi^A \, (A=1,\cdots,8)$ is the 
standard fermionic superpartner of 
$X^I$. These are  2-component Majorana spinors which also transform under 
$8_s$ spinor representation of $SO(8)$. While spinors $\chi$ and $\zeta$ 
make the supersymmetric 
partners of $\eta^I$ and $C^I_{(2)}$ respectively. 
Note 
that, in this formulation we have scalar-tensor action and the B-F 
(Chern-Simons) gauge actions at an 
equal footing. They are both topological in nature and only propagating 
fields are the matter fields $X^I$'s. 

With this action,  we obtain the following ${\cal N}=8$ supersymmetry 
variations for  the fields 
\footnote{In our 
convention $\gamma^\m_{\alpha\beta}$ are real and commute with 
$\Gamma^I_{A\dot A}$. The spinors 
$\bar\chi=\chi^T\gamma^0$, see Appendix for details.} 
\bea
&&\delta X^I= i \bar \epsilon 
\tilde\Gamma^I\psi,~~~~~\delta\psi=-(\not\!\partial X^I-\eta^I\not\!B) 
\Gamma^I\epsilon, \br
&& \delta A_\mu={i\over 2} \eta^I\bar\epsilon 
\gamma_\mu\tilde\Gamma^I\psi
-{i\over 2} X^I \bar\epsilon \gamma_\mu \tilde\Gamma^I\chi
,~~~~~\delta 
B_\mu=0 \br
&& \delta \eta^I= i\bar\epsilon \tilde\Gamma^I\chi, ~~~~~
\delta \chi=-\not\!\partial\eta^I \Gamma^I\epsilon,\br
&&\delta 
C^I_{\m\n}=
{i}\bar\epsilon \tilde\Gamma^I \gamma_{\mu\nu}\zeta
,~~~
\delta\zeta= (\not\!\partial(\not\!B X^I)+ {1\over 
2}\epsilon^{\mu\nu\lambda}\partial_\mu C_{\n\l}^I) \Gamma^I\epsilon
\eea
under which $U(1)$  action \eqn{MCS1a} 
remains invariant. 
The supersymmetry parameters $\epsilon^{\dot A}$ are eight 2-component 
real spinors  belonging to the
$8_c$ representation of $SO(8)$.

\subsection{The triviality of $U(1)$ }
It would be useful to verify that the Abelian case presented above is 
nothing but the rewriting of the 
non-interacting theory of  scalar fields 
describing the transverse motion of a  
membrane. For working this out, we first integrate out the auxiliary 
tensor field by 
using its equation of motion $\partial_\mu \eta^I=0$. So we substitute 
$\eta^I=g^I$ in the action. The action becomes
\bea\label{qMCS2}
&&S_{U(1)}=\int d^3x \, (
 -{1\ov2}(\partial^\mu X^I-B^\mu g^I )^2 
+{1\over2}\epsilon^{\mu\nu\lambda} B_\mu F_{\nu\lambda}
+ {i\over 2} \bar\psi \not\!\partial\psi)
\eea
Notice that now we have  the Stueckelberg invariance namely: 
$$\delta X^I=g^I 
f(x),~~\delta B_\mu=f(x). $$
We have two possibilities here either we integrate out $B_\mu$ or 
integrate 
out $A_\mu$ first. 

I)
Let us first take the case of integrating out the $A_\mu$ field. 
We presume that field strength $F_{\m\n}$ to be a 
fundamental field and impose its Bianchi identity by 
adding a Lagrange 
multiplier term ${1\over 2}\int \partial_\m\tau 
F_{\n\l}\epsilon^{\m\n\l}$. 
Here $\tau$ is periodic $\tau\sim \tau+1$. The 
Abelian action then becomes
\bea\label{qMCS1}
&&S_{U(1)}=\int d^3x \, (
 -{1\ov2}(\partial^\mu X^I-g^I B^\mu)^2 + {1\over 2}
\epsilon^{\mu\nu\lambda} (B_\mu+\partial_\m\tau) F_{\nu\lambda}
+ {i\over 2} \bar\psi  \not\!\partial\psi)
\eea
We then integrate out the $A_\mu$ which is auxiliary gauge field through 
the equation of motion
$B_\m+\partial_\m\tau=0$. 
\bea\label{qMCS3}
&&S_{U(1)}=\int d^3x \left( 
 -{1\ov2}(\partial^\mu X^I+g^I\partial_\mu\tau )^2 
+ {i\over 2} \bar\psi\not\!\partial\psi 
\right) 
\eea
Since $\tau$ transforms as 
$\tau\to\tau-\lambda$ under $B_\mu\to~B_\mu+\partial_\m\lambda$, using 
this freedom we can always gauge fix 
$\tau=0$. We 
are left with 
\bea\label{qMCS3v}
&&S_{U(1)}=\int d^3x \left( 
 -{1\ov2}(\partial^\mu X^I)^2 
+ {i\over 2} \bar\psi \not\!\partial \psi 
\right) 
\eea
which is nothing but the known non-interacting $SO(8)$ theory 
for  
single membrane and accounts for all the degrees of freedom. The $X^I$'s 
are the 
modes describing the transverse motion of a membrane on  $R^8$.  

II) The second option could have been that we integrate out $B_\mu$ field 
first by using the dNS 
equation. In which case $B_\mu$ field eats up one of the $X^I$'s through 
shift symmetry and it
becomes heavy which also breaks $SO(8)$ spontaneously. After substituting 
dNS 
equation we obtain the gauge 
action representing
a single D2-brane 
\bea\label{qMCS3v1}
&&S_{U(1)}=\int d^3x \left( 
 -{1\ov2}\sum_{i=1}^7(\partial^\mu X^i)^2 
-{1\over 4 g_{0}^2}F_{\m\n}F^{\m\n}+ {i\over 2} \bar\psi \not\! 
\partial\psi 
\right) 
\eea
and it has explicit $SO(7)$ invariance. It is obvious that both of these 
actions \eqn{qMCS3v} and \eqn{qMCS3v1} are equivalent in 3D.

\subsection{U(N) case} 
Now having studied the simpler Abelian case in the STBF formulation, we 
now 
set to determine the fermionic 
content of the non-Abelian 
action \eqn{MCS}. We find that the fermionic content in the action 
remains the same as in Abelian case except 
that  now $\psi^A$ is in the adjoint of $U(N)$, while 
the pair $(\chi^A,\zeta^A)$ 
remains gauge singlet. But there are also additional fermionic terms. The
full action can be written as 
\bea\label{un1}
S_{U(N)}&= & 
\int d^3x \bigg[ \Tr(
 -{1\ov2}(D^\mu X^I-\eta^I B^\mu)^2 
-U(\eta,X)- (B^\l X^I)\partial_\lambda\eta^I\! 
+ {1\ov2}\epsilon^{\mu\nu\lambda} B_\mu F_{\nu\lambda})
\br &&~~~~~~~ -{1\ov2}\epsilon^{\mu\nu\lambda} C_{\mu\nu}^I 
\partial_\lambda\eta^I 
+ 
\Tr{i\over 2} \bar\psi \not\!D\psi +i\Tr \bar\chi (\not\!B \psi)
-i \bar \chi \zeta\br && ~~~~~~~
-\Tr{i\over 2} \bar\psi \Gamma_{IJ}\eta^I[X^J,\psi] 
-\Tr{i\over 2} \bar\psi \Gamma_{IJ}[X^I,X^J]\chi  \bigg]
\eea
All the spinors transform under $8_s$ of the R-symmetry group $SO(8)$ as 
usual. The covariant fermionic derivative is given by
\be 
D_\mu\psi=\partial_\mu\psi-[A_\mu,\psi]\ .
\ee
All bosonic covariant 
derivatives in the action are usual gauge covariant 
derivatives involving $A_\m$. 

Note that, the  action \eqn{un1} has a shift (gauge) invariance
under which tensor fields also transform 
\bea\label{shift2}
&&\delta_1 
B_\mu=D_\mu f, ~~~
\delta_1 X^I=\eta^I f, ~~~
\delta_1 C_{\mu\nu}^I=\epsilon_{\mu\nu\lambda}\Tr\partial^\lambda (f 
X^I),\br
&&~~~\delta_1 \psi= f \chi, ~~~\delta_1 \zeta= -\Tr\not\!\partial( f\psi) 
\ .   
\eea
One will easily notice that $\delta_1 U=0$. That is the shifts $\delta_1 
X^I$  do not change the potential. In addition 
there is an usual $U(N)$ gauge symmetry 
involving $A_\mu$ fields as discussed in the review section.

Determining the supersymmetric variations for non-Abelian case is 
rather difficult.  But we know that our theory can be mapped into 
L-BLG, see the section (3.4) below, so the task becomes easier. 
We take the 
lead from 
 L-BLG work \cite{verlinde} and following the map in the 
section 
(3.4) we determine that
the supersymmetry variations for the  non-Abelian STBF are 
\bea\label{susyun}
&&\delta X^I= i \bar \epsilon 
\tilde\Gamma^I\psi,~~~\delta\psi=-(\not\!D X^I-\eta^I\not\!B) 
\Gamma^I\epsilon-{1\over 3!}V^{IJK}\Gamma_{IJK}\epsilon, \br
&& \delta A_\mu=
{i\over 2} \eta^I \bar\epsilon 
\gamma_\mu\tilde\Gamma^I\psi
-{i\over 2} X^I \bar\epsilon \gamma_\mu\tilde\Gamma^I\chi
,~~~~~\delta 
B_\mu=i\bar\epsilon\gamma_\mu\tilde\Gamma_I[X^I,\psi] \br
&& \delta \eta^I= i\bar\epsilon\tilde\Gamma^I\chi
, ~~~~\delta 
\chi=-\not\!\partial\eta^I \Gamma^I\epsilon,~~~\delta 
C^I_{\m\n}={i}\bar\epsilon\tilde\Gamma^I\gamma_{\m\n}\zeta
,~~~~
\br &&
\delta\zeta=
\Tr(\not\!\partial(\not\!B X^I)\G^I-{1\over2}(\not\!\partial 
X^I)X^{JK}\G^{IJK})\epsilon+ {1\over 
2}\epsilon^{\mu\nu\lambda}\partial_\mu C_{\n\l}^I \G^I\epsilon\ .
\eea
One can check that the straightforward reduction of 
\eqn{susyun} to the Abelian case gives 
the susy variations determined in the previous section. 
We note that at no stage did we
require to invoke a tri-algebra, as all expressions in the action, 
including the expressions like $V^{IJK}$ or  
$\Tr(\not\!\partial X^I)X^{JK}$ in 
the susy variations, do involve normal Lie-brackets. \footnote{ To 
make it clear that, although by looking at various
triple products one would like 
to believe that these terms may come from some hidden tri-algebra 
structure, but it is 
not immediately clear if this will be true while we are in STBF set 
up, $i.e.$  having tensor fields explicitly in the action. In order to 
realise 3-algebra explicitly 
we should first dualise or map STBF back to the L-BLG.}
As an important next step, we will now  show that the STBF theory 
can actually be 
mapped to the familiar L-BLG theory where  3-algebra structure 
becomes a favorable simplifying tool. 

\subsection{Generalised dNS relation, gauge fixing: Equivalence of STBF 
and BLG theory}
There has been  an expectation that the STBF theory constructed via tensor 
field inclusion method must be related to L-BLG tri-algebra theory 
somehow.\footnote{ I 
am grateful to Neil Lambert for raising this issue and for sharing his 
insight.}
Particulalrly the STBF action has  got a lot of similarity with the 
L-BLG action 
\cite{verlinde} involving the propagating ghost fields.   
Here we try to establish this missing equivalence between the STBF and 
the L-BLG. Let us separate the bosonic STBF Lagrangian in the following
manner
\be\label{fb0}
L_0(X^I,\eta^I,B_\m, A_\m) - 
{1\over 2}\epsilon^{\m\n\l}
 \partial_\m\eta^I C^I_{\nu\lambda}
\ee
where $L_0$ contains all the terms in STBF Lagrangian except  
the tensor fields. With out any loss of the content we can introduce {\it 
new} set of gauge fields 
$\hat 
A^I_\m$ (having mass dimension 
${1\ov2}$ and transforming in the $8_v$) through a total derivative 
term 
\be\label{fb1}
L_0(X^I,\eta^I,B_\m, A_\m) - 
{1\over 2}\epsilon^{\m\n\l}
 \partial_\m\eta^I (C^I_{\nu\lambda}-
2\partial_\n\hat A^I_\l)
\ee
These eight  gauge fields are the singlets of $U(N)$.
However, it is important 
to notice that these do not modify any of the equations obtained 
previously from STBF Lagrangian 
and actually these fields are just a kind of 
harmless spectator fields. However, due to these, 
action \eqn{fb1} has additional shift symmetry; 
$$\delta 
C^I_{(2)}=d\alpha^I_{(1)},~~~~\delta\hat 
A^I_{(1)}=\alpha^I_{(1)}. $$ 
This  invariance can  be utilised to 
gauge fix the gauge field 
$\hat A^I_\mu=0$. This is what we have considered throughout in the 
paper. The  BdGPT relation remains unchanged and it is 
\be
{\delta L_0\over \delta\eta^I}=-
{1\over 2!}\partial_\m 
C^I_{\nu\lambda}\epsilon^{\m\n\l}
\ee
To recall
the dNS duality
relation involving adjoint fields is
\be\label{eqxi0}
{1\over 2! }\epsilon^{\mu\nu\lambda} 
F_{\nu\lambda}
=(\eta^I B^\mu-D^\mu X^I)\eta^I+ X^I\partial^\mu\eta^I\ ,
\ee
which defines relationship between gauge fields $A_\mu,B_\mu$ and 
the scalars $X^I$ all in adjoint of the gauge group.
But it does not involve any tensor fields.

We now  wish to define a `generalised' dNS (gdNS)  duality 
relation involving only singlet fields, namely
\be\label{gdns}
{1\over 2}\epsilon^{\m\n\l} (
 C^I_{\nu\lambda}-\hat F_{\nu\l})
\equiv 
(\hat C^{I\m} -\partial^\m X^I_{-})
\ee
where field strength $\hat F^I\equiv d\hat A^I.$
Only difference in gdNS relation and the dNS 
equation 
is that the gdNS equation  
involves 2-rank tensor fields along with Abelian gauge fields $\hat 
C^I_\m,\hat 
A^I_\m$ and the scalar $X^I_{-}$. The $\hat A^I_\m$ need not explicitly 
appear in the action as it can be eaten up by the tensor field, while its 
presence only adds to total derivative terms in the action. Using this 
generalised  
duality relation, the  STBF action \eqn{fb1} 
can now be 
written in the L-BLG form
\be
L_0(X^I,\eta^I,B_\m, A_\m) - 
 \partial_\m   X^{I}_{+} (\hat C^{I \m}-
\partial^\m  X^I_{-})
\ee
where we redefined $\eta^I\equiv X^{I}_{+}$ for identification. This 
is the action  constructed in \cite{verlinde,bobby,Bandres:2008kj}. 
Since here both 
$X^{I}_{+},X^{I}_{-}$ are propagating fields with lightlike metric, due to 
this the 
L-BLG  action has ghost 
degrees of freedom, which are eliminated 
through the
gauge fixing as discussed in \cite{Bandres:2008kj,verlinde}. With  $SU(N)$
gauge symmetry the 
BLG theory has been shown to acquire a Lorentzian  
 tri-Lie-algebra structure \cite{verlinde}. 

So far that was for mapping the bosonic content on the two 
sides. The 
fermionic content 
is mapped as follows. Specifically, if the fields $C^I_{\m\n},~\hat 
C^I_\l,~ X^I_{-}$ are chosen to
have their fermionic partners given by  $\zeta,~\hat\chi,\psi_{-}$ 
respectively. Then the fermionic map from STBF to L-BLG is given by
\be\label{gdns1}
 \zeta=\hat\chi-\not\!\partial\psi_{-}\, .
\ee
All other fermions remain unchanged under this map. We note that, 
we could make a gauge choice 
$\hat A^I_\m=0$, likewise we can also  
have a choice where we can set $X^I_{-}=0=\psi_{-}$ in L-BLG, see 
\cite{Bandres:2008kj}.  

\section{Supersymmetric 
Vacua}

The moduli space of vacua in the STBF theory is larger than the 
3D super Yang-Mills theory. Our main aim is to determine vacua which will 
have nontrivial tensor backgrounds.
 
The first set of solutions are the constant 
$X^I$ configurations where $B_\m$ and $A_\m$ 
fields are vanishing \cite{verlinde,hs}. 
So for these solutions  $D_\mu X^I-\eta^IB_\mu=0$.
 If we take  
 $\eta^I=g^I$ and $ C_{\m\n}^I$ being constants in the vacuum, we 
only require
\be
X^{IJ}=[X^I,X^J]=0.
\ee
 That means $X^I$'s must be commuting 
(diagonal) $N\times N$ matrices. 
It gives the moduli space to be exactly that of $N$  M2-branes 
on flat $R^8$.   
Since for these solutions the 
 $V_{IJK}$ and $(D_\mu X^I-\eta^I B_\mu)$ are vanishing thus
all supersymmetric
fermionic variations altogether vanish.  
So these make the maximally supersymmetric solutions of STBF theory.
These STBF vacua are the same  as those of L-BLG  \cite{verlinde} 
and it is 
consistent with the map discussed in section (3.4).  
We comment that for any finite coupling $(g^I)^2$ the theory actually 
describes the super Yang-Mills theory of D2-branes, the membrane 
theory is obtained 
only in the strong coupling limit of it as elaborated in \cite{verlinde}. 

\noindent{\bf Noncommuting solutions:}

I) An interesting  case  arises when
 $C^I_{(2)}$ is taken to be nontrivial. For this let 
 us take the tensor components  to be dependent on the 
spatial coordinates 
\be
d C_{(2)}^I= m^I(x) dx^0\wedge dx^1\wedge dx^2\ ,
\ee
 where
$m^I(x)$ is a function which we shall determine next.  We  
still take
$A_\m=0$ and first discuss the case with $B_\mu=0$. Following from  
$\eta^I$ 
equation of motion, we find 
$X^I$ and $m^I$ will be related via   
\be\label{new4}
 {1\over2!} g^{[I} \Tr( X^{JK]}X_{JK}) =m^I
\ee
This ought to describe  a noncommuting ({\it fuzzy}) 
configuration  of 
membranes. We further simplify to the special 
case where $(\eta^8=g_{YM}, \eta^i=0)$ and $(m^8=m(x),~m^i=0)$.
The $X^I$ equations of motion reduce to
\bea
&& \partial_\mu X^8=0 \br
&&\partial_\mu\partial^\mu X^i + (g_{YM})^2 [X^{ij}, X^j]=0
\eea
and following from \eqn{new4}, $X^i$'s are  to satisfy the 
constraint
\be\label{new4a}
 {1\over2} g_{YM} \Tr( X^{ij}X_{ij}) =m(x)
\ee
Thus $X^8$ has to be constant and the $X^i$ equations  are 
in fact satisfied 
by the Nahm equation
\be\label{nahm1}
\partial_\sigma X^i={1\over 2} g_{YM}\epsilon^{ijk} [X^j,X^k]
~~~(i=1,2,3)
\ee
where  $x_2\equiv \sigma$. The Nahm equation has a simple solution
\be\label{nahm2}
X^i={1\over g_{YM}\sigma}\Sigma^i 
\ee
where $\Sigma^i$ form an  $SU(2)$ subalgebra, 
$[\Sigma^i,\Sigma^j]=\epsilon^{ijk} \Sigma^k$, and are in the 
$N\times N$ representation. The  $X^i(\sigma)$ are well defined for 
$\sigma >0$. The  $\sigma=0$ is the location of 
the boundary brane. From eq.\eqn{new4a} we 
 determine  (for large N)
\be
m(\sigma)\propto {N^3 \over g_{YM}^3 \sigma^4} .
\ee
It suggests  that the 0-1 component of the tensor field falls off as 
\be
C_{01}\propto ({ N\over g_{YM}\sigma})^3 \ 
.
\ee
Thus we have seen that the nontrivial tensor fields 
can be present in the STBF theory for a noncommuting 
fuzzy sphere configuration. It can be interpreted as an indication of the 
presence of the boundary M5-brane. 
 Actually 
from M5-brane point of view $\sigma$ is a transverse coordinate 
and $\sigma\to 0$ is like probing the region where boundary branes are 
located. 

We can  express
\be
[X^i,X^j]={1\over g_{YM}\sigma} \epsilon^{ijk}X^k \ .
\ee
The physical radius square of the sphere (for large $N$) at a fixed 
location $\sigma$ is
\be
r(\sigma)=\sqrt{{1\over N}\Tr (\sum_i X_i^2)}\sim { N \over 
2 g_{YM}\sigma}.  
\ee
So the radius of the sphere varies with the location, $\sigma$, and it 
blows 
up near the boundary $\sigma=0$, the location of M5-brane. 
These fuzzy sphere solutions have earlier been 
obtained in D1-D3 
system  \cite{myers} and  BMN-matrix model \cite{bmn}. 

II)
Although in the above we have taken  $B_\mu=0, ~X^8=constt$ while 
solving for the fuzzy solution, instead 
we can  take $B_\mu$ to be pure gauge such 
that it
solves  $\partial_\sigma X^8-g_{YM}B_\sigma=0$.
The 
fuzzy-sphere configuration above is still a solution with
\be\label{new4b}
 {1\over2} g_{YM} \Tr( X^{ij}X_{ij}) =\partial^\l\Tr(B_\l 
X^8)={1\over g_{YM}}\Tr\partial_\sigma( 
X^8\partial_\sigma X^8) 
\ee
but with a constant tensor field. 
We then determine that 
\be
X^8\sim {1\over\sqrt{3}} \sum_{i=1}^3 X^i\ .
\ee

These two fuzzy-sphere solutions, one with nontrivial tensor field 
and 
the other with a constant or vanishing value, are not related via 
infinitesimal (shift) symmetry as discussed  in eq.\eqn{shift2}. 
Hence we  conclude that  there can be a nontrivial tensor field 
background for the fuzzy 
sphere solution in the membrane B-F theory. 
 
We also 
check that for the sphere solution all fermionic variations \eqn{susyun}
can be made to vanish 
identically. Note that, from 
$\delta_s\psi=0$ the  
arbitrary  spinors need to
satisfy
\be
~~~\gamma^2_{\beta\alpha}\G^{1238}_{\dot B\dot 
A}\epsilon^{\dot A\alpha}=-\epsilon^{\dot B \beta}
\label{kl1}
\ee
where $\G^{1238}_{\dot B\dot 
A}$ is a lower diagonal
component of  $16\times 16$ matrix
\be
\bar\G_p=\bar\G_1\bar\G_2\bar\G_3\bar\G_8=\left( \begin{array}{cc}  
\tilde\G^{1238}_{B A}& 0\cr 
0& \G^{1238}_{\dot B \dot A}\end{array}\right) \ee
In our conventions 
$\G^{1238}_{\dot B\dot 
A}=-\sigma_2\times {\bf 1}_2 \times \sigma_2$ and  the property that 
$(\bar 
\G_p)^2=1$. Corresponding to \eqn{kl1} the 32 component Weyl spinors 
$\hat\epsilon=(0,\epsilon)$
would then satisfy\footnote{ Here $\hat\gamma^\m=\gamma^\m \times 
\G^{ch}$, 
with $\G^{ch}=diag(-{\bf 1}_{8},{\bf 1}_8)$ being the chirality operator. 
While 
$\hat\G^{1238}={\bf 1}_2\times \bar \G_p$.}
\be
~~~\hat\gamma^2\hat\G^{1238}\hat\epsilon=-\hat\epsilon
\label{kl2}
\ee
Thus  $ \hat\g^2\hat\G^{1238}$  will act as a projector for 
an arbitrary  Weyl spinor $\hat\epsilon$. We 
can always choose the eigenvalues such that
the 8 components of the spinors remain intact. Likewise other 
fermionic variations also identically vanish. Thus
we find that fuzzy sphere is a $1/2$-supersymmetric solution of 
STBF 
theory. This is in agreement with the other known cases of fuzzy 2-sphere
in the literature \cite{myers,bmn}. Previous works on fuzzy sphere 
solutions in 
Bagger-Lambert-Gustavsson membrane theory can be found in 
\cite{sphere1,sphere2}.

 \section{Conclusion}
We have discussed the supersymmetrisation of the membrane B-F theory 
(STBF) having 
dual-pairs of non-propagating  scalar and  tensor fields. The construction 
has been  based on ordinary 
 Lie-algebra structure. This construction leads us to  a  
supermembrane theory with tensor fields
which has  $U(N)$ gauge symmetry, $SO(8)$ R-invariance as well as 
the scale invariance. There are no free parameters in the 
action as those can be scaled away. The 
theory  does not have propagating ghost degrees of 
freedom as the tensor fields are topological in nature. However, we 
do find that a Lorentzian tri-algebra 
structure can emerge if the tensor fields are dualised into propagating 
scalar  fields 
via a `generalised' dNS like duality and as a consequence the theory goes 
over to the known L-BLG formulation \cite{verlinde}. We have explicitly 
shown that there exists 
a fuzzy $S^2$ solution which supports a nontrivial 2-rank tensor field and 
it is 
 $1/2$-supersymmetric. It will be interesting if we can find a fuzzy $S^3$ 
solution in STBF theory.

\vskip.5cm
\noindent{\it Acknowledgements:}\\
 I am thankful to  Anirban Basu and Bobby Ezhuthachan for  
useful discussions. I am also grateful to Neil Lambert for the careful 
reading of an earlier draft and the helpful comments. 
Finally I  wish to thank the organisors of the Indian Strings Meeting 
 ISM'08, Pondicherry for providing an exciting environment during the 
workshop.

\appendix{

\section{Conventions:}

The 3D Clifford algebra with signature $(-++)$  is 
given by
\be
\{\gamma^\mu,\gamma^\nu\}=2\eta^{\m\n}\ .
\ee
We choose a real representation where $\gamma^0=i\tau^2$, 
$\gamma^1=\tau^1$
and $\gamma^2=\tau^3$. The $\tau$'s are the Pauli matrices. 
The  matrix $C=\gamma^0$ and satisfies  
$C\gamma^\mu C^{-1}=-(\gamma^\mu)^T$. The fermionic invariants can be 
constructed involving 2-component Majorana spinors as 
$$\bar\chi\zeta=\chi^T\gamma^0\zeta=\bar\zeta\chi,~~~ 
\bar\chi\gamma^\mu\zeta=-\bar\zeta\gamma^\mu\chi,~~~
\bar\chi\gamma^{\mu\nu}\zeta=-\bar\zeta\gamma^{\mu\nu}\chi,\cdots\ 
.$$
The $\gamma$-commutators are defined as 
\bea
&& \gamma^{\m\n}={1\over 
2}[\gamma^\mu,\gamma^\nu]=\epsilon^{\m\n\l}\gamma_\lambda\ 
, ~~~\gamma^{\mu\nu\lambda}=\epsilon^{\m\n\l} {\bf 1}
\eea
while the Levi-Civita tensor is $\epsilon^{012}=1$.

For the internal space the $SO(8)$ Dirac algebra requires 
$16\times 16$ 
reducible matrices, 
$$\bar\G^I=\left( \begin{array}{cc} 0 & \G^I_{A\dot A}\cr 
(\tilde\G^I)_{\dot B B}&0\end{array}\right) $$
 corresponding to the 16-component 
Majorana spinors 
$$\Psi=(\psi^A_s,\psi^{\dot A}_c)$$
which are formed from the Weyl spinors $\psi^A_s$ and $\psi^{\dot A}_c$. 
We denoted $\tilde\Gamma^I=(\Gamma^I)^T$ and the spinorial
indices are
$A,\dot A=1,\cdots,8$.  
The $\bar\G^I$'s satisfy the algebra
\be
\{\bar\G^I,\bar\G^J\}=2\delta^{IJ}
\ee
provided $\G^I_{A\dot A}$ satisfy the  relations
\bea 
&&\G^I_{A\dot A}\tilde\G^J_{\dot A B}
+\G^J_{A\dot A}\tilde\G^I_{\dot A B}=2\delta^{IJ}\delta_{AB}\br
&&\tilde\G^I_{\dot A A}\G^J_{A \dot B}
+\tilde\G^J_{\dot A A}\G^I_{ A\dot B}=2\delta^{IJ}\delta_{\dot A\dot B}.
\eea
Similarly, we can also define antisymmetric products
\bea 
&&\G^I_{A\dot A}\tilde\G^J_{\dot A B}
-\G^J_{A\dot A}\tilde\G^I_{\dot A B}=2\G^{IJ}_{AB}\br
&&\tilde\G^I_{\dot A A}\G^J_{A \dot B}
-\tilde\G^J_{\dot A A}\G^I_{ A\dot B}=2\G^{IJ}_{\dot A\dot B}.
\eea

The component matrices $\G^I_{A\dot A}$ can also be treated as real 
Clebsch-Gordon 
coefficients. With these $\Gamma$'s an
$SO(8)$ invariant quantity can be constructed by combining  
 the vector and  two fermionic 
representations. We shall be using the real representation
\bea
&&\G_1=\sigma^2\times \sigma^2\times\sigma^2, \br
&&\G_2=1\times \tau^1\times\sigma^2, \br
&&\G_3=1\times \tau^3\times\sigma^2, \br
&&\G_4=\sigma^2\times 1\times\tau^1 \br
&&\G_5=\sigma^2\times 1\times\tau^3 \br
&&\G_6=\tau^1\times\sigma^2\times 1 \br
&&\G_7=\tau^3\times\sigma^2\times 1 \br
&&\G_8=1 \times 1\times 1 \ ,
\eea
where $\sigma^2=i\tau^2$.
See for more details 
on the representations  in
\cite{gsw}.

\section{The topological scalar-tensor 3D action}
Let us  discuss here a topological scalar-tensor (ST) action just like we 
have B-F 
(Chern-Simons) gauge action in 3D. We can write 
\bea\label{nhd}
 S_{ST}\sim\int d^3x  (-{1\ov2}\epsilon^{\mu\nu\lambda} 
C_{\mu\nu}^I\partial_\lambda\eta^I  -i 
\bar \chi \zeta)
\eea
The fermions 
$\chi^A$ and $\zeta^A$ make the supersymmetric partners for  $\eta^I$ and 
$C^I_{\m\n}$ and belong to $8_s$ representation of $SO(8)$.
 All the fields in the Lagrangian are nonpropagating 
(auxiliary) fields. 
The equations of motion of the fermions are simply $$\zeta=0=\bar\chi\ .$$ 
There is an obvious gauge invariance under $\delta 
C^I_{\m\n}=\partial_{[\m} \lambda^I_{\n]}$.

The action \eqn{nhd} does possess ${\cal N}=8$ supersymmetry 
under the infinitesimal variations 
\bea
&& \delta \eta^I= i\bar\epsilon\tilde\Gamma^I\chi, ~~~\delta 
\chi=-\not\!\partial\eta^I \Gamma^I\epsilon,\br
&&\delta 
C^I_{\m\n}={i}\bar\epsilon\tilde\Gamma^I\gamma_{\mu\nu}\zeta,~~~
\delta\zeta= {1\over 
2}\gamma^{\m\n\l}\partial_{\mu}C_{\nu\lambda}^I\Gamma^I\epsilon\ .
\eea
The supersymmetry parameters $\epsilon^{\dot A}$ are 8 two-component 
Majorana spinors  belonging to $8_c$ representation of $SO(8)$. 

The above  scalar-tensor action can be constructed in analogy with 
Chern-Simons gauge action \cite{schwarz}
\bea\label{nhd1}
 S=-\int d^3x  ({1\ov2}\epsilon^{\mu\nu\lambda} 
A_{\mu}F_{\nu\lambda} + i 
\bar \chi \chi) .
\eea
which is topological in nature.

\section{Romans' type IIA supergravity: The BdGPT duality}
The massive type IIA maximal supergravity \cite{romans} in 10 dimensions 
is known to have 
a cosmological constant term proportional to $m^2$ alongwith mass terms 
for 2-rank tensor fields. Due to this the action does not 
have the known
$Z_2$ invariance of type II strings, under which RR $p$-form potentials 
flip their sign, unless the mass parameter simultaneously changes its 
sign. It was 
 interestingly
suggested in \cite{berg} that the 
mass parameter $m$ could, in fact, be lifted to a 0-form,  $F_{(0)}$, 
which 
in turn 
can be Hodge-dualised to a 10-form field strength 
\be
F_{(10)}\equiv dA_{(9)}\ .
\ee
Particularly, the 
D8-branes are charged under 9-form potential $A_{(9)}$, which are  1/2-BPS 
solutions of the theory. Under this 
`localisation' of the Romans' mass, the action goes over to \cite{berg}
\be 
L_m^{IIA}(A_{(p)};m) \to
L_m^{IIA}(A_{(p)};F_{(0)}) + F_{(0)}\wedge dA_{(9)}
\ee
In 
fact, the 9-form potential $A_{(9)}$ plays the role of a Lagrange 
multiplier 
field. It imposes the constraint that 
\be
dF_{(0)}=0,
\ee
 i.e. in the vacuum $<F_{(0)}>=m$. The 
 duality relation between $F_{(0)}$ and $F_{(10)}$ is nothing but the 
$F_{(0)}$ 
equation of motion
\be
{\delta L_m^{IIA}(A_{(p)};F_{(0)})\over\delta F_{(0)}}=-\ast_{10} 
dA_{(9)}\ .
\ee 
We  call this as Bergshoeff-de-Roo-Green-Papadopoulos-Townsend 
(BdGPT) duality relation. 
In this formulation the massive type IIA SUGRA regains the $Z_2$ symmetry 
under 
which the local fields transform as 
\be
F_{(0)}\to -F_{(0)}~~~~  {\rm and}~~~~ A_{(9)}\to -A_{(9)}  .
\ee
instead of a constant mass parameter $m\to -m$.
 It is this similar argument which we incorporated
 in constructing an $SO(8)$ invariant B-F theory \cite{hs}.  
In that construction there are eight constant couplings $g^I$ which 
transform under $SO(8)$. However, the theory transforms along with the 
couplings. So it was needed to make these couplings localised, 
$$g^I \mapsto 
\eta^I(x)\ .$$ 

\vskip1cm
\centerline{----------------------------}
}


\end{document}